\documentclass[preprint, authoryear]{elsarticle}

\usepackage{natbib}
\usepackage[utf8]{inputenc}
\usepackage[T1]{fontenc}
\usepackage{lmodern}
\usepackage[fleqn]{mathtools}
\usepackage{rotating}
\usepackage{amssymb}
\usepackage{amsmath}
\usepackage{graphicx}
\usepackage{caption}
\usepackage{epstopdf}
\usepackage{stackrel}
\usepackage{soul}
\usepackage{color}
\usepackage{tabularx}

\newcommand*\diff{\mathop{}\!\mathrm{d}}

\renewcommand*{\underline}{\ul}

\newdefinition{rmk}{Remark}
\newproof{pf}{Proof}
\newproof{pot}{Proof of Theorem \ref{thm2}}

\newtheorem{theorem}{Theorem}[section]

\begin{document}

\title{On Capital Allocation under Information Constraints\\[5pt]
{\small(First Version: 14.06.2019, Current Version: 20.04.2020)}}

\author[hhu]{Christoph J.~B\"orner}
\ead{Christoph.Boerner@hhu.de}
\author[hhu]{Ingo Hoffmann\corref{cor1}}
\ead{Ingo.Hoffmann@hhu.de}
\author[hhu]{Fabian Poetter}
\ead{Fabian.Poetter@hhu.de}
\author[hhu]{Tim Schmitz}
\ead{Tim.Schmitz@hhu.de}


\cortext[cor1]{Corresponding author. Tel. : +49 211 81-15258; Fax. : +49 211 81-15316}

\address[hhu]{Financial Services, Faculty of Business Administration and Economics, \\ Heinrich Heine University D\"usseldorf, 40225 D\"usseldorf,
Germany}

\begin{abstract}
Attempts to allocate capital across a selection of different investments are often hampered by the fact that investors' decisions are made under limited information (no historical return data) and during an extremely limited timeframe. Nevertheless, in some cases, rational investors with a certain level of experience are able to ordinally rank investment alternatives through relative assessments of the probabilities that investments will be successful. However, to apply traditional portfolio optimization models, analysts must use historical (or simulated/expected) return data as the basis for their calculations. 
This paper develops an alternative portfolio optimization framework that is able to handle this kind of information (given by an ordinal ranking of investment alternatives) and to calculate an optimal capital allocation based on a Cobb-Douglas function, which we call the Sorted Weighted Portfolio (SWP). Considering risk-neutral investors, we show that the results of this portfolio optimization model usually outperform the output generated by the (intuitive) Equally Weighted Portfolio (EWP) of different investment alternatives, which is the result of optimization when one is unable to incorporate additional data (the ordinal ranking of the alternatives). To further extend this work, we show that our model can also address risk-averse investors to capture correlation effects.
\end{abstract}

\begin{keyword}
Capital Allocation \sep Cobb-Douglas Utility Function \sep Decision Theory \sep Uncertainty \sep Portfolio Selection Theory.\\[10pt]
\textit{ORCID IDs}: 0000-0001-5722-3086 (Christoph J.~B\"orner), 0000-0001-7575-5537 (Ingo Hoffmann), 0000-0003-4588-7728 (Fabian Poetter), 0000-0001-9002-5129 (Tim Schmitz).
\end{keyword}

\maketitle

\thispagestyle{empty} 
\section*{Key Messages}
\begin{itemize}
	\item Capital allocation with no available historical risk/return data
	\item Portfolio theory under the restriction of an extremely limited timeframe
	\item Investors are able to ordinally rank different alternatives based on their experience
	\item Sorting and appropriate allocation leads to dominant results compared to the EWP
\end{itemize}
\section{Introduction}
Decision-makers, e.g., financial investors, often face alternatives that do not differ at first glance. This may be due to the availability of too little or too much information.\\

According to the Laplace criterion, the distribution of future outcomes should be considered the same for all these alternatives. From a Bayesian perspective, this reflects the prior distribution (\cite{Bayes1763}). A risk-neutral decision-maker would choose any of the alternatives. For risk-averse decision-makers, the stochastic dependency of the alternatives would be important. However, decision-makers often have their own information at their disposal, e.g.,\ from experience. If we assume that this information is well founded, it can be interpreted as a signal that updates the a priori probability to the posterior probability. This signal does not lead to a decision under certainty but to a new distribution for each alternative. The posterior distribution is then the rational basis for decision-making.\\

This paper builds on the concept of Bayes' theorem but models a specific decision-making situation that has not yet been researched. The key point here is that the decision-maker has relevant information that is available only in a particular, limited way. The decision-maker's knowledge thus makes it possible to ordinally rank the alternatives with respect to the assessment of their relative contributions to overall success.\\

However, the basic consideration in this context is that investors often have to make decisions quickly and without statistical reflection. In the context of Bayes' theorem, this means that they cannot determine the type I or type II error of their information (or the conditional probabilities). Therefore, we consider how rational decisions are made when the experience of decision-makers leads to relevant but incomplete information.\\

The abovementioned problem arises in many different financial and nonfinancial use cases. For example, consider situations in which rational investors have to make decisions about possible investments very quickly and using a limited range of information, such as (so-called) elevator pitches, venture capital (VC) investment pitches or other auction-like situations, e.g., \ specialized investment fairs or auctions of livestock or art. Such situations seem typical for VC investors\footnote{See \cite{Rind1981} for elaboration of the typical business model of venture capitalists.}, who considers to invest in a new business idea or an expanding start-up. Generally those situations are accompanied by a comparably high level of uncertainty about future outcomes (\cite{Rind1981}). VC fund managers have to choose among alternatives by allocating their money across a number of entrepreneurs of varying talent (e.g., gold miners with differently located claims). These examples are from the area of finance, but examples from other areas are conceivable. Thus, we assume that the concept discussed below is always applicable in situations in which allocation decisions have to be made by using a limited range of information.\\

Our research topic combines three different strands of the literature: (i) modern portfolio theory (asset management), (ii) decision theory and (iii) production theory. Modern portfolio theory is based on the work of \cite{markowitz52,markowitz59}, who develops a portfolio optimization framework for given (historical/simulated/expected) return data (\cite{elton17}). More recent works on portfolio optimization have sought to overcome the main shortcomings of Markowitz's traditional mean-variance approach. For example, \cite{Rockafellar2000} develop a Mean Conditional Value at Risk (CVaR) optimization framework to overcome the traditional assumption of normally distributed returns in the mean-variance framework.
In contrast to existing portfolio optimization models, we assume the presence of information constraints resulting from, e.g., historical track records that are too short (usually in the VC business) or time pressure. In other words, we address situations that occur in reality whenever financial investment decisions cannot be based on long data series (historical/expected/simulated returns), and therefore, a subjectivistic Bayesian formulation of probabilities becomes important for rational decision-making. We assume that a representative (rational) VC fund manager is able to create a (rational) ordinal ordering of investment alternatives based on experience (with previous investments) and industry knowledge. However, existing portfolio optimization models cannot cope with ordinal rankings\footnote{Ordinal rankings are traditionally used in the context of utility rankings in household theory in microeconomics (e.g. \cite{HicksAllen1934}). However, in portfolio theory, there are no existing portfolio optimization models that use ordinally ranked variables as input factors.} (alternative kinds of information). As a consequence, we develop an innovative optimization model based on a Cobb-Douglas production function, which is also applicable to decision situations with this kind of informational restriction and time pressure. In this sense, the article also offers advice for practice. To more concretely explain the key points of our optimization strategy, we refer to the capital allocation of the abovementioned VC fund manager across different start-ups as an illustrative use case below. In making this decision, investors not only rely on the allegedly best alternative but also allocate their money across all the investment alternatives. 
To the best of our knowledge, this is the first paper that extends traditional portfolio selection theory \cite{markowitz52,markowitz59} to a model that assumes limited information (an ordinal ranking of the investment alternatives).\\

In this paper, we show how an investor can improve his utility by considering an ordinal ranking of the alternatives. Considering the case of risk-neutral investors, we show
that portfolio optimization based on order statistics, which come into play as the ranking of the investment alternatives, outperforms the output generated by the (intuitive) Equally Weighted Portfolio (EWP) of investment alternatives, which would be the strategy of an investor who is unable or unwilling to use (additional) ranking information. We provide instructions on how to behave in situations with low information quality that contrast with the traditional portfolio optimization framework developed by Markowitz. To best address the work of Markowitz, we present an extension that modifies the assumptions about the investor's risk attitude. We show that our model is also able to address risk-averse investors, and hence, we are able to capture correlation effects.\\

The remainder of this paper is structured as follows:
Sec.\ \ref{2} introduces the abovementioned use case and develops a suitable optimization model for this capital allocation problem. Next, we focus on the illustrative solution of the special case of allocating VC between two (or more) start-up companies for risk-neutral investors who maximize their (monetary) output. Moreover, in Sec.\ \ref{PortfolioSelection}, we change the risk attitude of investors and assume risk aversion in the optimization model. The last section discusses the results and summarizes the key points.

\section{Portfolio Selection for Risk-Neutral Investors}\label{2}
\subsection{Methodological Framework}\label{ModelSolution}
Consider a use case in which a rational VC fund manager has the opportunity to invest capital $c_0$ in $n$ start-up companies that are operated by different (more or less successful) entrepreneurs. These are smaller individual companies, perhaps all from a specific industry, for which the respective entrepreneur needs external funding from an investor. Suitably scaled, the companies $i = 1, \ldots, n$ are assumed to have the same possible maximum absolute output $a$, which is comparable to the absolute output potential of a start-up in a certain industry. The individually realized monetary output $c_{1,i}$ is also dependent on the entrepreneur's individual success (among other factors), which is modeled by the individual success factor $x_i$. This individual success factor is a stochastic variable, which is scaled between 0\% and 100\% and follows a standard uniform distribution: $x_i \sim {\cal U}(0,1)$. Thus, the abovementioned individual success factor reflects the relative potential of an individual start-up compared to its peer group (industry) and signals the extent to which the possible absolute maximum output $a$ can be exploited.\\
Hardly any other valid information is available to the investor. However, a rational investor can sort the companies based on the information available --- here, the (rational) estimated ranking of success factors. Using his or her experience, the investor decides that the output of company $i$ must be greater than or equal to that of company $j$ (by assumption). Hence, $0\leq ... \leq x_{j} \leq x_{i} \leq ... \leq 1$. In our simplified model, this ordinal ranking is based on the investor's evaluation of the entrepreneur's ability and success, as described by factor $x$, leading to total monetary output $c_1$.
This gives rise to the question of how the investor should distribute capital $c_0 = \sum_{i=1}^{n}c_{0,i}$ among the considered companies to maximize the expected total (monetary) output. Specifically, what amount of cash $c_{0,i}$ should be invested in company $i$?\\
To answer this question, we need to develop a methodological framework that can handle all the relevant information from the above-mentioned use case.
First, we assume that the investor's benefit from a company $i$ depends only on the combination of two factors: 1.\ the financial commitment to the company $c_{0,i}$ and 2.\ the individual success factor $x_i$. 
In addition, we assume that the resulting (monetary) benefit $c_{1,i}$ depends on the absolute amounts of both input factors. In our example, it seems intuitively obvious that the benefit the investor can derive from the potential of a successful entrepreneur increases due to the initial seed investment. In other words, a successful entrepreneur who can use a relatively large amount of seed money will generate a higher output than an identical entrepreneur with lower capital provision. Moreover, investors do not benefit from an entrepreneur's high individual success level if they do not invest any capital in the company in question ($c_{0,i} = 0$). With respect to this relationship, the (monetary) benefit as output in toto would be a function of the multiplication of the input factors. Furthermore, the contribution of each input factor to total output is determined by an individual partial elasticity.\\

To model such interdependency, the literature employs the Cobb-Douglas functional form. Charles W. Cobb and Paul H. Douglas formulated a production theory in the early $20^{th}$ century by combining the input factors of labor and capital to explain overall economic output (\cite{cobb28}). Their approach can be appropriately applied in the context considered in this paper. 
In the application below, the monetary output $c_{1,i}$ of company $i$ is defined by the following function:
\begin{flalign}\label{CDfkt}
c_{1,i} & = a' x_i^{\nu} c_{0,i}^{1-\nu},
\end{flalign}
where $a' > 0$ is a scale parameter with an appropriate unit of measurement to express the output value in a desired unit. $\nu$ denotes the partial elasticity of the introduced success factor $x_i$. The value $\nu$ is constant and influenced by the available technology. The model in Eq.\ \eqref{CDfkt} arises under the assumption that the sum of the elasticities of factors $x_i$ and $c_{0,i}$ equals 1.\\

We extend the original model to harmonize the Cobb-Douglas function with our application, which is characterized by a stochastic input factor representing entrepreneurial success. Hence, the input factor $x$ is a realization of the random variable $X$ assumed to be uniformly distributed between 0 and 1: $x \sim {\cal U}(0,1)$.\footnote{Assume that the success (here, output) of a venture is normally influenced by many different (observable and unobservable) factors. In addition to the size of the capital stock, factors include the quality of human resources; vulnerability to financial, operational, logistical or environmental risks; and efficiency issues. In this paper, we assume that the stochastic variable $X$ pools all of these factors to simplify the comparison of different companies.} 
Note that the random variable $X$ --- and internal company processes --- is not under the investor's control; instead, the investor chooses only the amount of seed money to finance the entrepreneur.\\

Our simple model in Eq.\ (\ref{CDfkt}) can then obviously describe situations in which a  realization $x$ of a random input factor $X$ is given, and the total output of money is a result of stochastics.
Recall that no time-series data are available to mitigate the influence of stochastics, which means formulating a deterministic model. This should apply in practice to the vast majority of cases --- particularly investing in start-up projects, where the key figures required for the investment decision (e.g., data about the balance sheet, the feasibility of the planned technology, the market demand and the historical track records of the founders) are often given only sparsely, reflecting information constraints in the investment process.\\

Now, let us consider not just one business but a number $i = 1,\ldots, n$ of companies for which the extended model in Eq.\ (\ref{CDfkt}) holds. Then, let $X_1, X_2, \ldots, X_n$ be a sample of random variables with realizations $x_i \sim {\cal U}(0,1)$ assumed to be independent and identically distributed. While all $X_i$ are random, uncontrollable quantities of the enterprise itself, the $c_{0,i}$ are deterministic, predeterminable quantities. The input factor $c_{0,i}$ can describe the amount of capital that an investor invests in company $i$. 
In this framework, we consider a budget constraint: $c_0  = \sum_{i=1}^{n} c_{0,i}$. Because the sum of the investor’s individual capital provisions can be scaled by $c_0$, we can rewrite our model in terms of portfolio weights (the traditional notation is $\omega_i$) in the asset management context. This leads to $\omega_i = \frac{c_{0,i}}{c_0}$ and the re-formulated budget constraint $1  = \sum_{i=1}^{n} \omega_i$. \\

To describe the individual (monetary) output $c_{0,i}$ of enterprise $i$, we can re-write Eq.\ (\ref{CDfkt}) and replace $c_{0,i}$ with $\omega_i$, which leads to
\begin{flalign}\label{CDfktNR}
c_{1,i} & = a x_{i}^{\nu} \omega_i^{1-\nu},
\end{flalign}
where we have considered a slight transformation of the constant $a'$, leading to a new constant $a = a' c_0^{1-\nu}$ in our model. As described above, the constant $a$ represents the absolute maximum output that can be gained by start-ups in the specific industry. Thus, the total (monetary) output of all the investments can be defined as follows:
\begin{flalign}\label{TotalBenefitR}
c_1 = a \sum_{i=1}^{n} x_{i}^{\nu} \omega_i^{1-\nu}.
\end{flalign}
Note that because $x_{i}$ is a random number, the individual (monetary) output $c_{1,i}$ for a company $i$ and the total (monetary) output $c_1$ are also random numbers.\\

As a starting point for the subsequent portfolio optimization, we define
\begin{flalign}\label{UtilityFunction}
\text{U} = \text{E}[c_1] - \frac{1}{2}b*\text{Var}[c_1]
\end{flalign}
as a (generalized) utility function of the investor, where $\text{E}[c_1]$ is the expected value of the random variable $c_1$, $\text{Var}[c_1]$ is the respective variance, and $b$ is a risk-aversion parameter. This utility function is the foundation of all portfolio optimization frameworks, which are run afterwards. \\

Before we can calibrate and run such portfolio optimization models, we need to distinguish two cases for a rational investor: in one, the investments are identical and neither is preferred (Case 1); in the second, the investor can sort the investment objects with respect to the factor $x$ (Case 2).\\

\noindent\underline{Case 1:}\\[6pt]
If the information available for the investment objects is very scarce, then in many cases, there is little option but to regard the objects as equivalent. Thus, a realization $x_1, x_2, \ldots, x_n $ of the random variables cannot be sorted by size.

We assume that the investor cannot influence the level of $x$. The only input factor that the investor can influence is the portfolio weight $\omega_i$ of a single investment object $i$. The question is how to choose the respective portfolio weights to allocate capital among the $i=1,\ldots, n$ possible investment objects such that the expected total (monetary) output is maximized.\\

Assuming that risk-neutral investors rank all investment alternatives with the same expected (monetary) output equally (ignoring the risk of $c_{1,i}$), we need to set $b = 0$ in the utility function of Eq.\ \eqref{UtilityFunction} and obtain
\begin{flalign}\label{UtilityFunctionSpecial}
\text{U} = \text{E}[c_1]. 
\end{flalign}
Consequently, the level of the expected (monetary) output is the only decision variable for risk-neutral investors. This leads to the following optimization problem for the portfolio weights $\omega_i$ of the individual investments:
\begin{flalign}\label{OptProbRandom0}
\max_{\{ \omega_1, \omega_2, \ldots, \omega_n\}} \textrm{U} \qquad \textrm{subject to }\qquad 0  = 1 - \sum_{i=1}^{n} \omega_i
\end{flalign}
in general and
\begin{flalign}\label{OptProbRandom}
\max_{\{ \omega_1, \omega_2, \ldots, \omega_n\}} \textrm{E}[c_1] \qquad \textrm{subject to }\qquad 0  = 1 - \sum_{i=1}^{n} \omega_i
\end{flalign}
under the assumption of risk-neutral investors ($b=0$). Thus, we must maximize the Lagrange function
\begin{flalign}\label{LagrangeRandom} \nonumber
{\cal L} & =  \textrm{E}[c_1] +\lambda\Big( 1 - \sum_{i=1}^{n} \omega_i\Big) \\ \nonumber
& = \sum_{i=1}^{n} \textrm{E}[c_{1,i}] +\lambda\Big( 1 - \sum_{i=1}^{n} \omega_i\Big) \\
& = a \sum_{i=1}^{n} \textrm{E}[x_{i}^\nu]\; \omega_i^{1-\nu} 
	+ \lambda\Big( 1 - \sum_{i=1}^{n} \omega_i\Big)
\end{flalign}
with respect to $\omega_i$, where $\lambda$ is the so-called Lagrange multiplier.\\

The expected value $\textrm{E}[x_{i}^\nu]$ is determined with respect to the uniform distribution and leads to $\textrm{E}[x_{i}^\nu] = \frac{1}{1+\nu}$.
Thus, the expected total money output is simply
\begin{flalign}\label{ExpValueURandom}
\textrm{E}[c_1] & =  \frac{a}{1+\nu} \; \sum_{i=1}^{n} \omega_i^{1-\nu}.
\end{flalign}
Finally, the Lagrange function in Eq.\ (\ref{LagrangeRandom}) becomes
\begin{flalign}\label{LagrangeNewRandom} 
{\cal L} 
& = \frac{a}{1+\nu} \; \sum_{i=1}^{n} \omega_i^{1-\nu} 
	+ \lambda\Big( 1 - \sum_{i=1}^{n} \omega_i\Big).
\end{flalign}
Typically, the optimization problem is determined by the partial derivatives of $\cal L$ and solving the system of equations. Note that the second derivative of $\cal L$ with respect to $\omega_i$ is negative and leads to a negative definite Hessian matrix if $\omega_i>0$ and $\nu>0$. Then, the optimal solution
\begin{flalign}\label{LSolutionRandom}
\omega_i^* & = \frac{1}{n} 
\end{flalign}
for $i=1, \ldots, n$ describes the maximum $\textrm{E}[c_1]$ in Eq.\ (\ref{ExpValueURandom}) depending only on the number of companies $n$. This solution is equal to the traditional EWP.\\

Substituting Eq.\ (\ref{LSolutionRandom}) in Eq.\ (\ref{ExpValueURandom}) provides the maximum total money output for Case 1:\footnote{A special degenerate case arises when $\nu = 0$ is used. In the present model, this is not an ecologically sensible choice. This eliminates the first input factor in Eq.\ (\ref{CDfktNR}). By doing so, Eq.\ (\ref{TotalBenefitR}) describes perfect substitutes, meaning that the investor can invest all his capital in any single object and achieve the maximum expected total money output: $\max \textrm{E}[c_1] = a$. In other words, if the individual absolute output $x_i$ no longer influences the investor’s money output, the individual weightings of the investment alternatives also become irrelevant to the investor because from his perspective, all alternatives are equal in their contribution to the total money output.}
\begin{flalign}\label{MaxURandom}\nonumber
{\cal B}_1 
& = \max_{\{ \omega_1, \omega_2, \ldots, \omega_n\}} \textrm{E}[c_1] \\
& = \frac{a}{1+\nu} n^\nu.
\end{flalign}

\noindent\underline{Case 2:}\\[6pt]
We now consider the case in which a rational investor can sort the investment objects based on the little information available. The investor still cannot influence the random factors $x$. Based on his assessment, however, the investor can derive a pairwise forecast of which investment has a higher $x$. This leads to an ordinal ranking of the eligible investment objects. Thus, a realization $x_1, x_2, \ldots, x_n $ of the random variables can be sorted by size, and we can conduct further analysis within the framework of so-called order statistics (\cite{kendall76}):
\begin{flalign}
0 \leq x_{(1)} \leq x_{(2)} \leq \ldots \leq x_{(n)} \leq 1.
\end{flalign}
\begin{figure}[h]
	\centering
	\captionsetup{labelfont = bf}
	\includegraphics[width=0.975\textwidth]{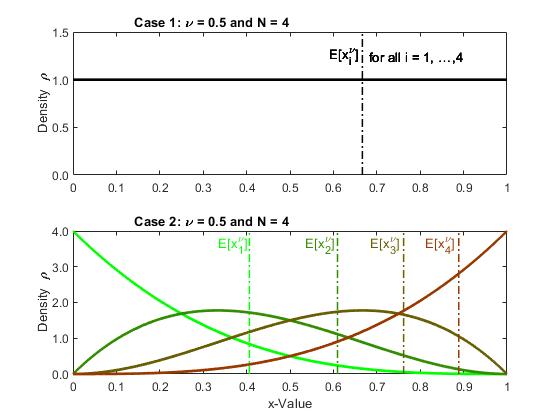}
	\begin{quote}
		\caption[Sample-Size 100]{\label{DiggersDensity} The influence of ordering the investment alternatives by their success factors on their individual densities}
		\label{fig:1}
	\end{quote}
\end{figure}
Applying those ordering statistics in this context leads to a shift in the individual density functions of the individual success factors $x_i$ (see Fig.\ \ref{DiggersDensity}). In contrast to Case 1, the investor has now an incentive to not allocate his or her capital equally across the different investment alternatives and find an alternative asset allocation, which reflects the updated information situation. We refer to this new approach as the Sorted Weighted Portfolio (SWP).\\ 

A new portfolio optimization approach is needed in this case. To describe the investors' monetary output $c_{1,i}$ of the investment in alternative $i$, we can once again re-write Eq.\ (\ref{CDfkt}) in terms of portfolio weights, which leads to
\begin{flalign}\label{CDfktN}
c_{1,i} & = a x_{(i)}^{\nu} \omega_i^{1-\nu}.
\end{flalign}
Thus, the total monetary output of all investments is again a random variable and given by $c_1 = \sum_{i=1}^{n} c_{1,i}$.
Note that $c_{1,i}$ also depends on $\omega_i$ and therefore generally has no sorting.\\

A question similar to that in Case 1 arises here: how should the input factors $\omega_i$ be chosen such that the expected total monetary output $\textrm{E}[c_1]$ of the investor is maximized (\cite{vonNeumann1953})? As a starting point, we have only sparse information: a limited budget, uniformly distributed $x$ and an ordinal ranking.\\

This situation leads to the same optimization problem as in Eq.\ (\ref{OptProbRandom}), except that in the Lagrange function of Eq.\ (\ref{LagrangeRandom}), another expectation value must be considered. 
The expected value $\textrm{E}[x_{(i)}^\nu]$ is now determined with respect to the order statistics. Since the distribution of the $i$th-order statistic $x_{(i)}$ in a random sample of size $n$ from the uniform distribution over the interval $[0, 1]$ is a beta distribution of the first kind with the following probability density (\cite{kendall76}):
\begin{flalign}\label{OrdStatPDF}
\rho(x; i, n) = \frac{1}{B(i, n-i+1)} x^{i-1}(1-x)^{n-i}
\end{flalign}
The expectation value can easily be calculated as follows:\\
\begin{flalign}\label{ExpValue}\nonumber
\textrm{E}[x_{(i)}^\nu] & = \int_{0 }^{1} x^\nu \rho(x; i,n) \diff x\\ \nonumber
& = \frac{\int_0^1 x^{i+\nu -1}(1-x)^{n-i} \diff x }{B(i, n-i+1)} \\
& = \frac{B(i+\nu, n-i+1) }{B(i, n-i+1)},
\end{flalign}
with $B(\alpha,\beta)$ being the beta function, which is also called a Euler integral of the first kind (\cite{abramowitz14}). With the discrete probability distribution
\begin{flalign}\label{DPDFHB}
p_i(\nu) & = \frac{1+\nu}{n} \frac{B(i+\nu, n-i+1) }{B(i, n-i+1)}
\end{flalign}
for $i=1,\ldots, n$ with integer $n>0$ and real $\nu>-1$ (\cite{hoffmann18}, Corollary 4), the expectation value can be rewritten:
\begin{flalign}\label{ExpValueD}
\textrm{E}[x_{(i)}^\nu] & = \frac{n}{1+\nu}\; p_i(\nu).
\end{flalign}
Therefore, the expectation of the total monetary output is simply
\begin{flalign}\label{ExpValueU}
\textrm{E}[c_1] & = a \; \frac{n}{1+\nu} \; \sum_{i=1}^{n} p_i(\nu) \; \omega_i^{1-\nu}.
\end{flalign}
The sum thus corresponds to the expected value of the transformed weights $\omega_i$ of the investment alternatives. Hence, under the assumption of the same utility function as in Eq.\ \eqref{UtilityFunctionSpecial} (still the special case of risk neutrality), the Lagrange function for Case 2 becomes
\begin{flalign}\label{LagrangeNew} 
{\cal L} 
& = a \; \frac{n}{1+\nu} \; \sum_{i=1}^{n} p_i(\nu) \; \omega_i^{1-\nu} 
	+ \lambda\Big( 1 - \sum_{i=1}^{n} \omega_i\Big).
\end{flalign}
As for Eq.\ (\ref{LagrangeNewRandom}), the optimization problem is determined by the partial derivatives of $\cal L$ and solving the system of equations. Note, as in Case 1, the second derivative of $\cal L$ with respect to $\omega_i$ is negative and leads to a negative definite Hessian matrix if $\omega_i>0$ and $\nu>0$. Then, the optimal solutions
\begin{flalign}\label{LSolution}
\omega_i^* & = \frac{p_i(\nu)^{\frac{1}{\nu}}}{\sum_{j=1}^{n} p_j(\nu)^{\frac{1}{\nu}}} 
\end{flalign}
for $i = 1, \ldots, n$ describe a maximum of $\textrm{E}[c_1]$ in Eq.\ (\ref{ExpValueU}) depending only on the elasticity $\nu$ and the number of companies $n$.\\

Substituting Eq.\ (\ref{LSolution}) into Eq.\ (\ref{ExpValueU}) provides the maximum total monetary output for Case 2:
\begin{flalign}\label{MaxU}\nonumber
{\cal B}_2 
& = \max_{\{ \omega_1, \omega_2, \ldots, \omega_n\}} \textrm{E}[c_1] \\ \nonumber
& = a \; \frac{n}{1+\nu} \;\left(\sum_{i=1}^{n} p_i(\nu)^{\frac{1}{\nu}}\right)^{\nu} \\
& = a \; \frac{n}{1+\nu} \;\big\| \boldsymbol{p}(\nu) \big\|_{\frac{1}{\nu}},
\end{flalign}
where $\big\| \boldsymbol{p}(\nu) \big\|_{\frac{1}{\nu}}$ denotes for fixed $\nu\in[0,1] $ the $\frac{1}{\nu}$-norm of the probability vector $\boldsymbol{p}(\nu) = (p_1(\nu), \ldots, p_n(\nu))$.\\

Comparing Case 1 and Case 2, the following theorem can be proven.
\begin{theorem}\label{theorem1}
\begin{flalign}\label{theoremEq1}
		{\cal B}_2 \geq {\cal B}_1
	\end{flalign}
for $n>1$ and $\nu\in[0, 1]\subset\mathbb{R}$. Equality applies if $\nu = 0, 1$.
\end{theorem}
\begin{pf}
With Eq.\ (\ref{MaxURandom}) and (\ref{MaxU}) Eq.\ (\ref{theoremEq1}) becomes
\begin{flalign}\label{inequality}
	n^{1-\nu}\big\| \boldsymbol{p}(\nu) \big\|_{\frac{1}{\nu}} \geq 1.
	\end{flalign}
Eq.\ (\ref{inequality}) follows directly from the Hölder inequality (\cite{abramowitz14}):
\begin{flalign}\label{hoelder}
	\sum_{k=1}^{n} |a_kb_k| 
	& \leq \Big(\sum_{k=1}^{n} |a_k|^u\Big)^{\frac{1}{u}} \Big(\sum_{k=1}^{n} |b_k|^v\Big)^{\frac{1}{v}}
	\end{flalign}
when $a_k = p_k(\nu)$, $b_k = \frac{1}{n}$, $u = \frac{1}{\nu}$ and $v = \frac{1}{1-\nu}$. The properties of $\boldsymbol{p}(\nu)$ show that equality holds when $\nu = 0, 1$. \qed
\end{pf}
Theorem \ref{theorem1} is consistent with the expectation that the investors' total monetary output of their investments increases if more information can be used in the investment decision. In other words, the results of the SWP (Case 2) dominate the results of the EWP (Case 1), which does not reflect the updated information situation caused by ordering the alternatives.
\subsection{Applications}
\subsubsection{Special Case --- Two Investments}\label{TwoAlternatives}
\begin{figure}[t]
	\centering
	\captionsetup{labelfont = bf}
	\includegraphics[width=0.975\textwidth]{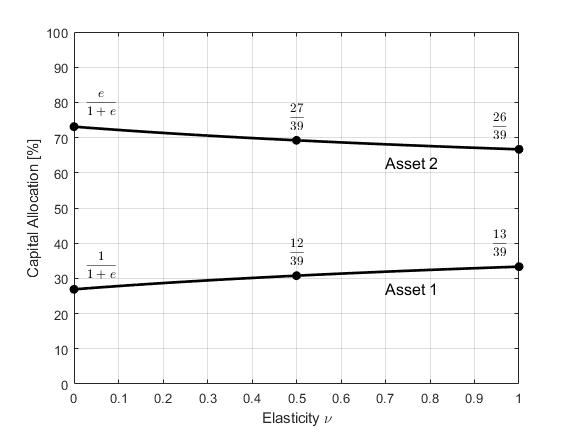}
	\begin{quote}
	\caption[Sample-Size 100]{\label{DiggersSVA} The dependence of the capital allocation on the elasticity $\nu$ considering two assets}
	\label{fig:1}
	\end{quote}
\end{figure}

As an example, consider a special of Case 2 in which seed money has to be distributed between two available investment objects. Hence, $n=2$ is the number of different investment objects, $\omega_1 + \omega_2 = 1$ is the budget constraint, and the only additional information assumed is $0 \leq x_{(1)} \leq x_{(2)} \leq 1$ for the success factor, with $x_{(i)}$ being the order statistic of a uniformly distributed random variable.
Thus, with Eq.\ (\ref{DPDFHB}),
\begin{flalign}\label{SimpleDPD}\nonumber
p_1(\nu) & =  \frac{1+\nu}{2} \; \frac{B(1+\nu, 2) }{B(1, 2)} = \frac{1}{2+\nu} \\ 
& \\[-8pt]
p_2(\nu) & =  \frac{1+\nu}{2} \; \frac{B(2+\nu, 2) }{B(2, 1)} = \frac{1+\nu}{2+\nu}. \nonumber
\end{flalign}
The optimal allocation of capital (SWP) is then described by the following proportions (cf.\ Eq.\ (\ref{LSolution})):
\begin{flalign}\label{SimpleSVA}\nonumber
\omega_1^* & =  \frac{1}{1 +(1+\nu)^{+\frac{1}{\nu}}}  \\ 
& \\[-8pt]
\omega_2^* & =  \frac{1}{1 +(1+\nu)^{-\frac{1}{\nu}}}.  \nonumber
\end{flalign}
Depending on the elasticity $\nu\in[0, 1]$, three special cases can be considered:
\begin{enumerate}[a)]
\item VC $\omega$ dominates the output money: $\nu \rightarrow 0$.
\item There is indifference between the input factors related to the output money:  $\nu = 0.5$.
\item The random number $x$ dominates the output money: $\nu \rightarrow 1$.
\end{enumerate}
Tab.\ \ref{SVA} reports the optimal capital allocation for these special cases, and Fig.\ \ref{DiggersSVA} depicts the allocation of capital between the two assets depending on the elasticity $\nu$.\\
\begin{table}[htbp]
	\captionsetup{labelsep=newline, justification=RaggedRight, singlelinecheck=false, labelfont = bf}
	\caption{Allocation of capital across $n=2$ investments for selected elasticities $\nu$}\label{SVA}
	\renewcommand{\arraystretch}{1.1}
	\begin{tabularx}{\textwidth}{  X  X  X }
		\hline
							& 						& 					\\[-6pt] 
		Elasticity 			& Asset 1               & Asset 2 			\\[1pt] 
		$\nu$					& $\omega_1$            & $\omega_2$		\\[6pt] 
		\hline
							& 						& 				  	\\[-6pt] 
 		$\nu \rightarrow 0$ & $\frac{1}{1+e}$       & $\frac{e}{1+e}$ 	\\[6pt]
		$\nu = 0.5$  		& $\frac{12}{39}$ 		& $\frac{27}{39}$ 	\\[6pt]
		$\nu \rightarrow 1$ & $\frac{13}{39}$       & $\frac{26}{39}$ 	\\[6pt] 
		\hline
	\end{tabularx}
\end{table}

Solving the allocation problem with unknown elasticity $\nu$: as a rule of thumb, capital $c_0=1$ can be distributed as seed money in a proportion $\frac{1}{3} : \frac{2}{3}$ between the two assets (usually, these are (minority) stakes in two early-stage companies) if no further information is known, except the ordinal ranking of the success factor ($ x_{(1)} \leq x_{(2)} $) and its assumed uniform distribution between 0\% and 100\%.

\subsubsection{Applications to More than Two Investments}
In the case of $n>2$ investments, (risk-neutral) investors face a more complex decision that is not covered by the rule of thumb developed above. Nevertheless, using the formula in Eq.\ (\ref{LSolution}), investors can easily calculate the optimal portfolio weights of the SWP for arbitrary $n$.\\

Considering the ordinal ranking $0 \leq x_{(1)} \leq x_{(2)} \leq \ldots \leq x_{(n)} \leq 1$, Tab.\ \ref{SVA_N} provides the optimal weights for $n = 3,4,5$ investments as an example.
\begin{table}[hbtp]
	\captionsetup{labelsep=newline, justification=RaggedRight, singlelinecheck=false, labelfont = bf}
	\caption{Sorted Weighted Portfolios considering $n=\{3,4,5\}$ investments and selected elasticities $\nu$}\label{SVA_N}
	\renewcommand{\arraystretch}{1.1}
	\begin{tabularx}{\textwidth}{Xcccccc}
		\hline
		& 			   &        &        &        &        &        \\[-6pt] 
		Investments & Elasticity & \multicolumn{5}{l}{Weights}          \\[1pt] 
		n          & $\nu$       & $\omega_1$ 
		& $\omega_2$  & $\omega_3$ & $\omega_4$ & $\omega_5$\\[6pt] \hline
		& 			   &        &        &        &        &        \\[-6pt] 
		3 & 0             & 0.1223 & 0.3315 & 0.5462 & -      & -      \\[6pt] 
		& $\frac{1}{4}$ & 0.1360 & 0.3321 & 0.5319 & -      & -      \\[6pt] 
		& $\frac{2}{4}$ & 0.1478 & 0.3326 & 0.5196 & -      & -      \\[6pt] 
		& $\frac{3}{4}$ & 0.1579 & 0.3330 & 0.5091 & -      & -      \\[6pt] 
		& 1             & 0.1667 & 0.3333 & 0.5000 & -      & -      \\[12pt] 
		4 & 0             & 0.0694 & 0.1881 & 0.3100 & 0.4325 & -      \\[6pt] 
		& $\frac{1}{4}$ & 0.0785 & 0.1917 & 0.3070 & 0.4228 & -      \\[6pt] 
		& $\frac{2}{4}$ & 0.0866 & 0.1948 & 0.3044 & 0.4143 & -      \\[6pt] 
 		& $\frac{3}{4}$ & 0.0937 & 0.1976 & 0.3021 & 0.4067 & -      \\[6pt] 
		& 1             & 0.1000 & 0.2000 & 0.3000 & 0.4000 & -      \\[12pt] 
		5 & 0             & 0.0446 & 0.1210 & 0.1993 & 0.2781 & 0.3570 \\[6pt] 
		& $\frac{1}{4}$ & 0.0510 & 0.1245 & 0.1995 & 0.2748 & 0.3502 \\[6pt] 
		& $\frac{2}{4}$ & 0.0568 & 0.1278 & 0.1997 & 0.2718 & 0.3440 \\[6pt] 
		& $\frac{3}{4}$ & 0.0620 & 0.1307 & 0.1998 & 0.2691 & 0.3384 \\[6pt] 
		& 1             & 0.0667 & 0.1333 & 0.2000 & 0.2667 & 0.3333 \\[6pt] \hline
	\end{tabularx}
\end{table}\\
Since the weights $\omega_i$ of the individual investments for fixed $n$ do not change substantially as the elasticity $\nu$ changes (see Tab.\ \ref{SVA_N}), we can focus on the limiting case $\nu \rightarrow 1$ in Eq.\ (\ref{LSolution}) and derive a rule of thumb for an unknown $\nu$:
\begin{flalign}\label{ruleofthumb}
\omega_i & = \frac{2 i}{n(n+1)}.
\end{flalign}
As a rough approximation, the capital allocation $\omega_i$ depends only on the position $i$ of the investment in the sorted alternatives and the total number $n$ of investment alternatives.
\section{Portfolio Selection for Risk-Averse Investors} \label{PortfolioSelection}
Modern portfolio theory describes situations in which an investor considers the return and risk of each investment when making a capital allocation. In addition to the return and the risk, the dependency structure of the investment opportunities in the optimization is used to construct an efficient portfolio (\cite{markowitz52}). Usually, these three characteristics are estimated from historical data (\cite{elton17}).\\

When allocating VC, we assumed in Sec.\ \ref{ModelSolution} that no historical market data are available to estimate the return and the risk. In the following, we show how the basic idea of modern portfolio theory can nevertheless be transferred to our use case --- the optimal allocation of VC.\\

An investor will consider not only the expected value $\textrm{E}[c_1]$ of the total monetary output but also the variance $\textrm{Var}[c_1]$ of the total monetary output (as a measure of risk) in the capital allocation process.
As we have seen, uncertainty over the total monetary output results only from the factor $x_i$ (random variable) of the individual total output function Eq.\ (\ref{CDfktNR}). We consider a constant elasticity $\nu$ for the input factor $x_i$. Thus, the random variable becomes $x_i^\nu$.\\

In Case 2 in Sec.\ \ref{ModelSolution}, a rational investor was able to sort the investment objects based on an assessment of the factors $x_i$. Case 2 is further examined below.\\

As a starting point, we calculate the variance of the total monetary output with respect to order statistic $x_{(i)}$, capital allocation $\omega_i$ for $i = 1, \ldots, n$ and $n$ investment objects:
\begin{flalign}\label{VarianzU} \nonumber
\textrm{Var}[c_1] & = \textrm{E}[c_1^2] - \textrm{E}[c_1]^2 \\ \nonumber
& = a^2 \sum_{i=1}^{n} \sum_{j=1}^{n} \Big\{ \textrm{E}[x_{(i)}^\nu x_{(j)}^\nu] - \textrm{E}[x_{(i)}^\nu]\textrm{E}[x_{(j)}^\nu] \Big\} \omega_i^{1-\nu}\omega_j^{1-\nu} \\ 
& = a^2 \sum_{i=1}^{n} \sum_{j=1}^{n} V_{ij} \; \omega_i^{1-\nu}\omega_j^{1-\nu}.
\end{flalign}
The calculation and representation of the covariance matrix $V_{ij} = V_{ij}(\nu, n)$ for $i,j = 1,\ldots,n$ is shown in \ref{covariancematrix} Eq.\ (\ref{EntireC}). Note that the order statistics are not necessarily stochastically independent; thus, we also have entries in the minor diagonals of the covariance matrix (in contrast to Case 1). As a consequence, we can show that the simple ordering of the investment alternatives leads not only to a shift in the densities of $x_i$ (as previously shown) but also to the existence of specific correlations between the investment alternatives (see the Appendix for example calculations), which form the foundation of diversification effects in modern portfolio theory. Similar to traditional portfolio selection theory, we use available information about the pairwise correlations to calculate the optimal portfolio (here, the SWP). By sorting the alternatives, the correlations of the alternatives now deviate from zero, which was the solution before obtaining additional information (the ordering), and become positive. This is relevant information for risk-averse investors because changes in the correlations also influence the portfolio risk (compared to the EWP).\\

Using the (generalized) objective function from Eq.\ \eqref{UtilityFunction} with $b > 0$, we define $\text{U} = \text{E}[c_1] - \frac{1}{2}b*\text{Var}[c_1]$. With the variance, the objective function to be maximized can now be specified for Case 2, and together with the budget function, the Lagrange function is simply:
\begin{flalign}\label{LagrangeKomplett}\nonumber
{\cal L} 
= & \quad \textrm{E}[c_1] - \frac{1}{2}b\;\textrm{Var}[c_1] 
	+ \lambda\Big( 1 - \sum_{i=1}^{n} \omega_i\Big) \\[6pt] \nonumber
= & \quad a \; \frac{n}{1+\nu} \; \sum_{i=1}^{n} p_i(\nu) \; \omega_i^{1-\nu} \\ \nonumber
  & \quad -\frac{1}{2} a^2 b \; \sum_{i=1}^{n} 
	  \sum_{j=1}^{n} V_{ij}\; \omega_i^{1-\nu}\omega_j^{1-\nu} \\ 
  & \quad +\lambda\Big( 1 - \sum_{i=1}^{n} \omega_i\Big).
\end{flalign}
As in Sec.\ \ref{ModelSolution}, $a>0$ is the absolute maximum output in the specific industry, and $b>0$ denotes an individual risk-aversion parameter.\\

The optimization problem is to find the maximum of the Lagrange function dependent on the capital allocation $\omega_i$ for $i=1,\ldots,n$:
\begin{flalign}\label{OptProb}
\max_{\{\omega_1, \omega_2, \ldots, \omega_n\}} {\cal L}.
\end{flalign}

The optimal solution $\omega_i^*$ depends on the number of investment objects $n$, the parameters $a$ and $b$ and the elasticity $\nu$. In a numerical determination of the optimal solution, the $\omega$-space should be limited to the positive orthant reflecting the long-only case, where no short positions are possible. Maximization of the Lagrange function Eq.\ (\ref{LagrangeKomplett}) thus takes place with respect to the inequality constraint $\omega_i>0$ for all $i$.\\

If the expected total monetary output $\textrm{E}[c_1]$ is ignored in the Lagrange function Eq.\ (\ref{LagrangeKomplett}) and only the variance $\textrm{Var}[c_1]$ of the total monetary output is considered in the objective function, then the optimal solution of the optimization problem describes the capital allocation in a minimum variance portfolio.

\section{Discussion and Conclusions}\label{DiscussionConclusion}
Allocating capital to a selection of different investment objects is often a problem when investors' decisions are made under limited information (no historical return data) and within an extremely limited timeframe. Nevertheless, in some cases, rational investors with a certain level of experience are able to ordinally rank these investment alternatives.\\

In this paper, we developed an innovative portfolio optimization framework that uses such ordinal rankings as the foundation for determining the optimal portfolio weights for certain investment alternatives. Assuming a risk-neutral investor, we provided a closed-firm solution for the optimal weight vector, which depends on the partial elasticity $\nu$.
For an unknown $\nu$, we developed a rule of thumb that capital $c_0 = 1$ should be distributed across the alternatives in a certain proportion, depending only on $n$. For $n=2$ investment opportunities and assessment $x_{(1)}<x_{(2)}$, we found the approximate distribution $\frac{1}{3}:\frac{2}{3}$ for capital. \\

We showed that in general, the SWP outperforms the intuitive EWP solution, which is traditionally the benchmark for portfolio optimization strategies in the literature and, in this special case, the result of optimization when it is not possible to account for additional information (an ordinal ranking of investment alternatives).\\

In the extension of the model to risk-averse investors, we were able to formulate $\textrm{E}[c_1]$ and $\textrm{Var}[c_1]$ as the classical input factors of Lagrangian optimization and to address correlation effects. However, in this case, there is no general algebraic closed-form solution available for the optimization model. Consequently, our model has important implications for practice and is a useful starting point for the development of additional extensions and practical applications in research.

\appendix
\section{Covariance Matrix}\label{covariancematrix}
In the following section, we determine the covariance matrix $V_{ij}(\nu, n)$. The covariance matrix is needed in Sec.\ \ref{PortfolioSelection} to calculate the variance of the total monetary output. Following Eq.\ (\ref{VarianzU}), we have
\begin{flalign}\label{VarianzUA}
V_{ij} (\nu, n) & =  \textrm{E}[x_{(i)}^\nu x_{(j)}^\nu] - \textrm{E}[x_{(i)}^\nu]\textrm{E}[x_{(j)}^\nu].
\end{flalign}
The last term of this equation can be calculated using Eq.\ (\ref{ExpValueD}).
Therefore, we only need to focus our attention on the first term. The evaluation of $ \textrm{E}[x_{(i)}^\nu x_{(j)}^\nu]$ leads to a matrix ${\bf M}$ with entries $M_{ij}$ and $i,j=1,\ldots,n$.\\

\noindent\underline{Case $j=i$:}\\[6pt]
If $j = i$, then $M_{ii}=\textrm{E}[x_{(i)}^\nu x_{(i)}^\nu] = \textrm{E}[x_{(i)}^{2\nu}] = \frac{n}{1+2\nu}\;  p_i(2\nu)$ (cf.\ Eq.\  (\ref{ExpValueD})).\\

\noindent\underline{Case $j>i$:}\\[6pt]
This case describes the situation in which the order statistic $x_{(i)}$ is smaller than the order statistic $x_{(j)}$. We first calculate the upper triangle of the matrix $\bf M$.
To determine the expected value $\textrm{E}[x_{(i)}^\nu x_{(j)}^\nu]$, the joint probability distribution of the order statistics is needed. The formulas become shorter if we write $u = x_{(i)} $ and $v= x_{(j)}$ with $u<v$. Then, the expectation value $\textrm{E}[u^\nu v^\nu]$ must be derived with respect to the joint probability distribution (\cite{kendall76})
\begin{flalign}\label{commondist}
\rho(u,v; i, j, n) = {\cal C}(n,i,j) \quad u^{i-1}(v-u)^{j-i-1}(1-v)^{n-j}
\end{flalign}
with the constant
\begin{flalign}\label{constdist}
{\cal C}(n,i,j) = \frac{\Gamma(n+1)}{\Gamma(i)\Gamma(j-i)\Gamma(n+1-j)}
\end{flalign}
and $\Gamma(\alpha)$ being the gamma function (\citet{abramowitz14}).
A double integration then provides the expectation value:
\begin{flalign}\label{ExpVal}\nonumber
& \textrm{E}[u^\nu v^\nu] \\[6pt] \nonumber
& = \int_{0}^{1}  \int_{0}^{v}  \; u^\nu v^\nu \rho(u,v; i, j, n) \diff u \diff v \\[6pt] \nonumber
& = {\cal C}(n,i,j) \int_{0}^{1}  \int_{0}^{v}  \; u^\nu v^\nu u^{i-1}(v-u)^{j-i-1}(1-v)^{n-j} \diff u \diff v. \\[6pt] \nonumber
& \hspace{0.5cm} \textrm{Substituting } w = 1 - v \textrm{ leads to }\\[6pt] \nonumber
& = {\cal C}(n,i,j) \int_{0}^{1}  \int_{0}^{1-w}  \; (1-w)^\nu u^{\nu+i-1}w^{n-j} (1-w-u)^{j-i-1} \diff u \diff w.\\[6pt] \nonumber
& \hspace{0.5cm} \textrm{Now, using the series representation, }  \\[6pt] \nonumber
& \hspace{0.5cm} \textrm{} (1-w)^\nu = \sum_{k=0}^{\infty} (-w)^k \frac{\Gamma(\nu+1)}{\Gamma(k+1)\Gamma(\nu+1-k)} \qquad \textrm{for } |w| < 1. \\[6pt] \nonumber
& \hspace{0.5cm} \textrm{ Since this is a convergent series with a finite limit, it follows} \\[6pt] \nonumber
& = {\cal C}(n,i,j) \sum_{k=0}^{\infty} (-1)^k \frac{\Gamma(\nu+1)}{\Gamma(k+1)\Gamma(\nu+1-k)} \times \\[6pt] 
& \hspace{1.5cm} \int_{0}^{1}  \int_{0}^{1-w}  \; u^{\nu+i-1}w^{n+k-j} (1-w-u)^{j-i-1} \diff u \diff w.
\end{flalign}
The last double integral is the representation of the two-dimensional beta function (\cite{waldron03}). Hence, the upper triangle of the matrix $\bf M$ is given by
\begin{flalign}\label{ExpVal2}\nonumber
 M_{ji}(\nu,n)& = \textrm{E}[u^\nu v^\nu]  =  \frac{\Gamma(n+1)\Gamma(\nu+1)\Gamma(\nu+i)}{\Gamma(i)\Gamma(n+1-j)} \times \\[6pt] 
 & \hspace{1.5cm}  \sum_{k=0}^{\infty} (-1)^k \frac{\Gamma(n+k+1-j)}{\Gamma(k+1)\Gamma(\nu+1-k) \Gamma(\nu+n+k+1)}. 
\end{flalign}

\noindent\underline{Case $j<i$:}\\[6pt]
For reasons of symmetry, $\textrm{E}[v^\nu u^\nu] = \textrm{E}[u^\nu v^\nu] $. Therefore, the lower triangle of the matrix is $M_{ij}(\nu,n) = M_{ji}(\nu,n)$.\\

\noindent Finally, the entire covariance matrix is given by
\begin{flalign}\label{EntireC}\nonumber
&\Big(V_{ij}(\nu, n)\Big)_{i,j = 1,\ldots,n}  \quad = \\[12pt]
&\left(
\begin{array}{cc}
\ddots & M_{ji} (\nu,n) - \left(\frac{n}{1+\nu}\right)^2\; p_j(\nu)p_i(\nu)\\[12pt]
\multicolumn{2}{c}{\left(\frac{n}{1+2\nu}\right)\; p_i(2\nu) - \left(\frac{n}{1+\nu}\right)^2\; p_i(\nu)^2}\\[12pt]
M_{ij} (\nu,n) - \left(\frac{n}{1+\nu}\right)^2\; p_i(\nu)p_j(\nu) & \ddots
\end{array}
\right)
\end{flalign}
Here, $M_{ji}(\nu,n)$ is defined in Eq.\ (\ref{ExpVal2}), and $p_i(\nu)$ is defined in Eq.\ (\ref{DPDFHB}).\\

\noindent By construction, for $n>1$, the covariance matrix is positive semi-definite. For $\nu \neq 0$, the random variables $x_{(i)}^\nu$ and $ x_{(j)}^\nu$ with $i\neq j$ in Eq.\ (\ref{VarianzUA}) are linearly independent, and the covariance matrix is positive definite and thus invertible (\cite{kendall76}).\\

With	
\begin{flalign}
\rho_{ij} = \frac{V_{ij}(\nu, n)}{\sqrt{V_{ii}(\nu, n)V_{jj}(\nu, n)}}
\end{flalign}
we can calculate the correlations between the sorted investment alternatives. For example, consider Tab.\ \ref{SVA_Correlation} for the case $n=4$ and different elasticities $\nu$. It can be observed that a considerable number of positive correlations are calculated when the alternatives are sorted. For elasticity $\nu = 0$, a numerical evaluation of the correlations is not possible. When the elasticity $\nu$ approaches 1, the correlations reach their maximums, and the correlation matrix becomes highly symmetrical. The correlations between alternatives are already large for a small elasticity $\nu$ and do not vary very much depending on $\nu$. Hence, if the elasticity $\nu$ is unknown, mere sorting leads to correlation effects.

\begin{table}[hbtp]
	\captionsetup{labelsep=newline, justification=RaggedRight, singlelinecheck=false, labelfont = bf}
	\caption{Correlation matrix for $n=4$ investments and selected elasticities $\nu$}\label{SVA_Correlation}
	\renewcommand{\arraystretch}{1.1}
	\begin{tabularx}{\textwidth}{Xl}
		\hline
		&                         \\[-6pt] 
		Elasticity $\nu$ & {Correlation matrix $\boldsymbol{\rho}$}        \\[6pt] \hline
		 			   &               \\
		 0.05          & $\begin{pmatrix}
							 1      & 0.5602 & 0.3578 & 0.2152 \\
							 0.5602 &    1   & 0.6438 & 0.3872 \\
							 0.3578	& 0.6438 &   1    & 0.6015 \\
							 0.2152	& 0.3872 & 0.6015 &   1    \\
						  \end{pmatrix}$  \\[36pt] 
		 0.25          & $\begin{pmatrix}
							 1      & 0.5844 & 0.3822 & 0.2318 \\
							 0.5844 &    1   & 0.6543 & 0.3967 \\
							 0.3822	& 0.6543 &   1    & 0.6063 \\
		 				 	 0.2318	& 0.3967 & 0.6063 &   1    \\
		 				  \end{pmatrix}$  \\[36pt] 
		 0.50          & $\begin{pmatrix}
		 					 1      & 0.6024 & 0.3988 & 0.2433 \\
		 					 0.6024 &    1   & 0.6620 & 0.4038 \\
		 					 0.3988	& 0.6620 &   1    & 0.6100 \\
		 					 0.2433	& 0.4038 & 0.6100 &   1    \\
		 				  \end{pmatrix}$  \\[36pt] 
		 0.75          & $\begin{pmatrix}
							 1      & 0.6103 & 0.4063 & 0.2486 \\
							 0.6103 &    1   & 0.6656 & 0.4073 \\
							 0.4063	& 0.6656 &   1    & 0.6118 \\
							 0.2486	& 0.4073 & 0.6118 &   1    \\
						  \end{pmatrix}$  \\[36pt] 
		 1.00          & $\begin{pmatrix}
							 1      & 0.6124 & 0.4082 & 0.2500 \\
							 0.6124 &    1   & 0.6667 & 0.4082 \\
							 0.4082	& 0.6667 &   1    & 0.6124 \\
							 0.2500	& 0.4082 & 0.6124 &   1    \\
						  \end{pmatrix}$  \\[24pt] \hline
	\end{tabularx}
\end{table}
\section*{Declaration of Interest}
The authors report no conflicts of interest. The authors alone are responsible for the content and writing of the paper.


\end{document}